\documentclass[twocolumn,nolinenumbers]{aastex631}

\usepackage{bm}
\usepackage{amsmath}
\shorttitle{Evidence of a polar circumbinary planet}
\shortauthors{R. G. Martin et al.}

\begin{document}

\title{AC Her: Evidence of the first polar circumbinary planet}

\author[0000-0003-2401-7168]{Rebecca G. Martin}
\affil{Nevada Center for Astrophysics, University of Nevada, Las Vegas,
4505 South Maryland Parkway, Las Vegas, NV 89154, USA}
\affil{Department of Physics and Astronomy, University of Nevada, Las Vegas,
4505 South Maryland Parkway, Las Vegas, NV 89154, USA}
\author[0000-0002-4636-7348]{Stephen H. Lubow}
\affiliation{Space Telescope Science Institute, 3700 San Martin Drive, Baltimore, MD 21218, USA}
\author[0000-0002-0543-6730]{David Vallet}
\affil{Nevada Center for Astrophysics, University of Nevada, Las Vegas,
4505 South Maryland Parkway, Las Vegas, NV 89154, USA}
\affil{Department of Physics and Astronomy, University of Nevada, Las Vegas,
4505 South Maryland Parkway, Las Vegas, NV 89154, USA}
\author[0000-0002-2208-6541]{Narsireddy Anugu}
\affiliation{The CHARA Array of Georgia State University, Mount Wilson Observatory, Mount Wilson, CA 91023, USA}
\author[0000-0001-8537-3583]{Douglas R. Gies}
\affiliation{Center for High Angular Resolution Astronomy and Department
of Physics and Astronomy, Georgia State University, P.O. Box 5060, Atlanta,
GA 30302-5060, USA}



\begin{abstract}
We examine the geometry of the post-asymptotic giant branch (AGB) star binary AC Her and its circumbinary disk. We show that the observations describe a binary orbit that is perpendicular to the disk with an angular momentum vector that is within $9^\circ$ of the binary eccentricity vector, meaning that the disk is close to a stable polar alignment.  The most likely explanation for the very large inner radius of the dust is a planet within the circumbinary disk.  This is therefore both the first reported detection of a polar circumbinary disk around a post-AGB binary and the first evidence of a polar circumbinary planet. We consider the dynamical
constraints on the circumbinary disk size and mass. 
The polar circumbinary disk feeds circumstellar disks with gas on orbits that are highly inclined  with respect to the binary orbit plane.
The resulting circumstellar disk inclination could be anywhere from coplanar to polar depending upon the competition between the mass accretion and  binary torques.
\end{abstract}

\keywords{accretion, accretion disks --- binaries: general --- hydrodynamics --- planets and satellites: formation.}


\section{Introduction}

Many post-asymptotic giant branch (AGB) stars have a disk that is Keplerian and stable \citep{deRuyter2006,Bujarrabal2013,Gallardo2021}. All post-AGB stars with a disk are in a binary and the disk is circumbinary  \citep{Bujarrabal2013,Izzard2018}. The disk may be formed in the wind of the parent star, during non-conservative mass transfer or common envelope ejection \citep[e.g.][]{Kashi2011,Izzard2023}. The binaries have orbital periods in the range $100-3000\,\rm day$ \citep{Oomen2018} and eccentricities up to about $0.6$, which may be a natural outcome of the stellar evolution process and binary-disk interactions \citep{Sepinsky2007,Sepinsky2007b,Dermine2013,Oomen2020, Zrake2021, Dorazio2021}.
The disks are remarkably similar to protoplanetary disks with masses  up to about $0.01\,\rm M_\odot$ and  sizes up to about $1000\,\rm au$ \citep{Sahai2011,Bujarrabal2015}. The accretion rate from the inner edge of the circumbinary disk is up to about $10^{-7}\,\rm M_\odot \, yr^{-1}$ \citep{Bujarrabal2018,Bollen2020}.

Around 10\% of the  post-AGB circumbinary disks have a lack of near infrared excess \citep{Kluska2022,Corporaal2023}, similar to transition disks around main-sequence stars.
AC Her \citep{VanWinckel1998,Gielen2007} is an example of a post-AGB binary system with a large inner hole in the dust. The observed binary and disk parameters of the system are shown in Table~\ref{table2}  that is based on values given in \cite{Hillen2015} and \cite{Anugu2023}.  The inner edge of the dust disk is about ten times the binary separation \citep{Hillen2015}, much farther out than the radius where tidal truncation from the binary would predict the inner disk edge \citep[e.g.][]{Artymowicz1994,Hirsh2020} and much farther out than the expected dust sublimation radius \citep{Kluska2019}. 
The best explanation for the large hole in the dust is the presence of a circumbinary planet \citep[e.g.][]{Kluska2022,Anugu2023} that creates a pressure maximum in the disk and traps dust grains while allowing gas to flow past the planet \citep[e.g.,][]{Pinilla2012,Pinilla2016,Zhu2013,Francis2020}. 

\begin{figure*}
    \centering  
        \includegraphics[width=0.4\columnwidth]{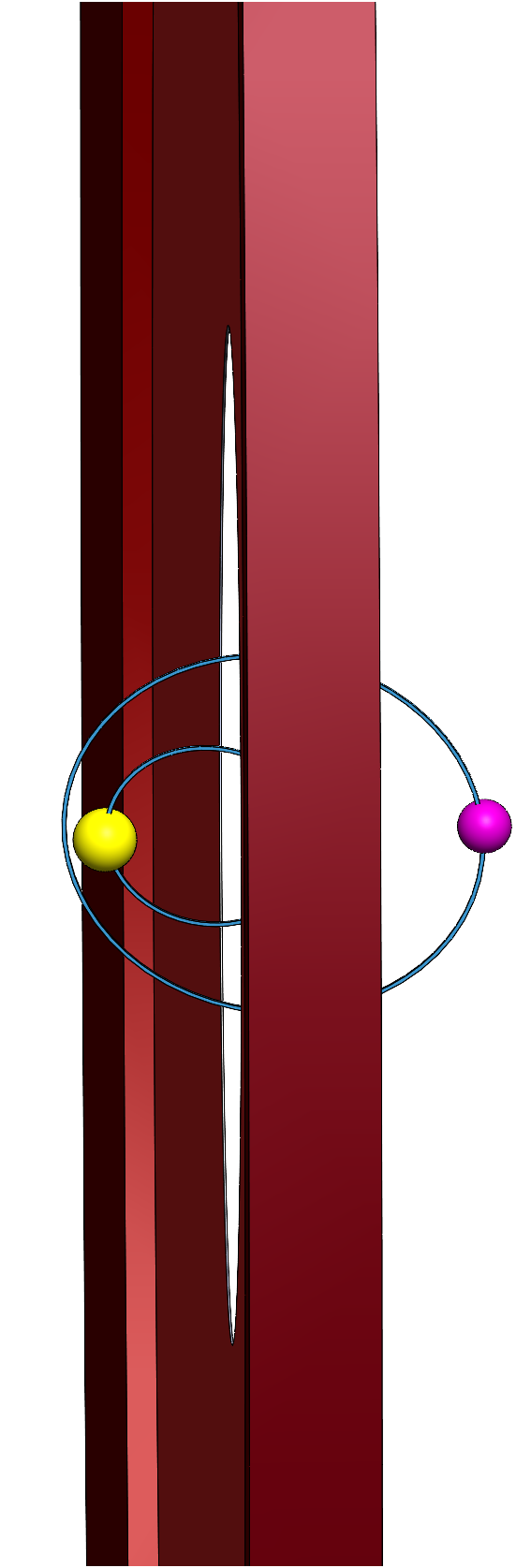}   
        \hspace{1cm}
        \includegraphics[width=1.4\columnwidth]{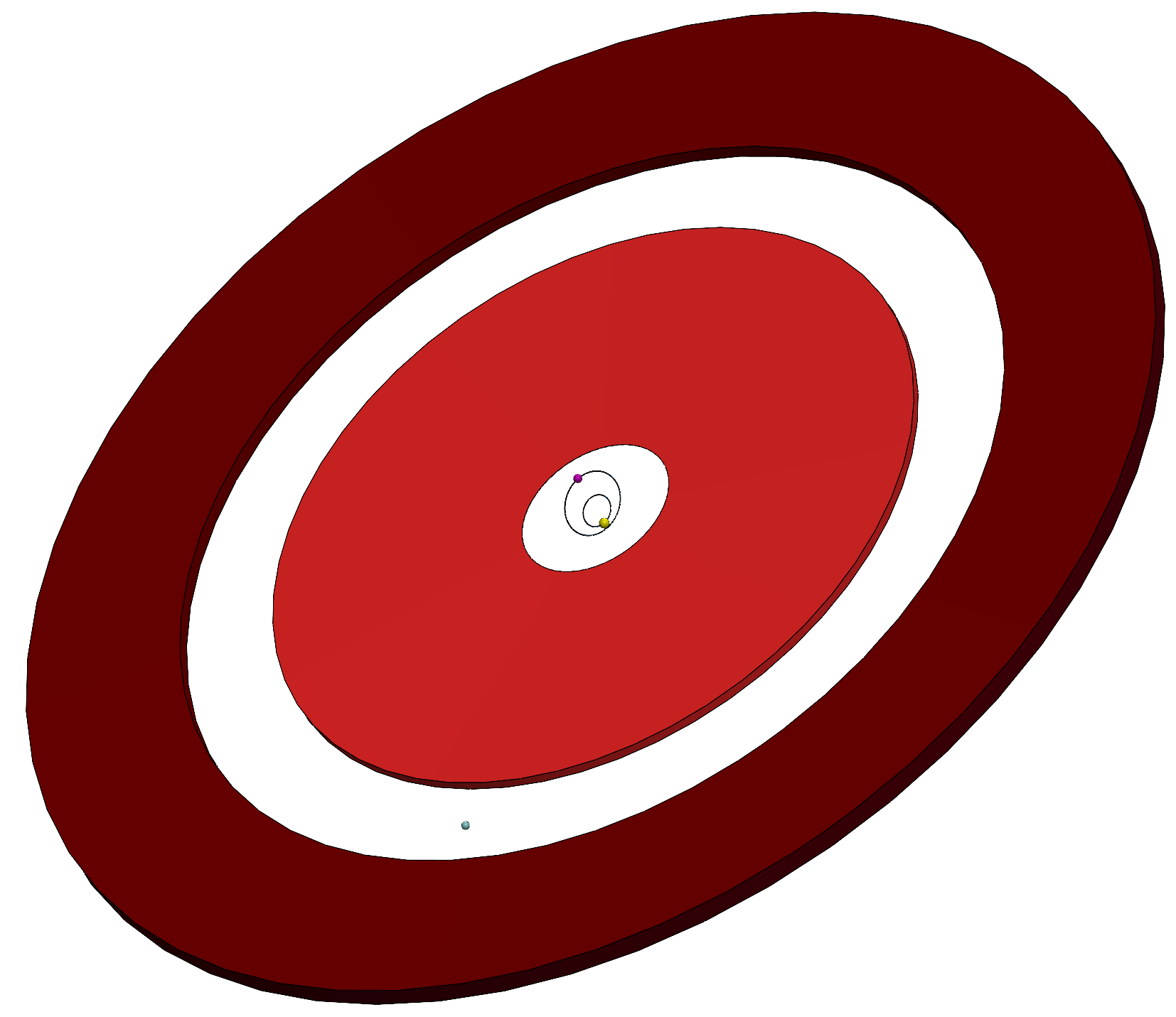}   
     \caption{3D visualisations showing two different views of the AC Her binary system with a polar circumbinary disk. The disk has two parts that are separated by the gap that the putative planet has cleared out. The orbital radius of the planet is shown at $8\,a_{\rm b}$ (Section~\ref{polarplanet}). The outer disk (dark red) is the gas and dust disk that is observed (Section~\ref{geo}).  The outer edge of the observed disk 
     extends beyond what is shown in the figure.
     The possible inner disk (light red) is a dust depleted disk that has not been observed (Section~\ref{polarplanet}).   The inner disk is shown to extend from $1.6\,a_{\rm b}$ to $7\,a_{\rm b}$.   The  post-AGB star (magenta sphere) and  the companion (yellow sphere) are shown at apastron separation. The binary orbits are to scale relative to each other but the sizes of the stars and the planet are not.     The left panel shows a view with the disk almost edge on and right panel shows the observed view in the plane of the sky. AC Her is located at a distance of $1402\,\rm pc$ \citep{Gaia2016}. The observationally inferred binary semi-major axis is $a_{\rm b}=2.01{\,\rm mas}=2.83\,\rm au$ \citep{Anugu2023}.  }
    \label{cartoon}
\end{figure*}

In Section~\ref{geo} we examine the geometry of the AC Her system and show that the circumbinary disk is highly misaligned to the binary orbit. In fact, the disk is close to alignment with the binary eccentricity vector, a configuration known as polar alignment (see the left panel of Fig.~\ref{cartoon}). This is a stable state for a circumbinary disk around an eccentric binary \citep{Aly2015,Martin2017,Lubow2018,Zanazzi2018,Cuello2019,Smallwood2020,Rabago2023}. Previously, there have been observations of two polar aligned circumbinary gas disks \citep{Kennedy2019,Kenworthy2022} and one debris disk \citep{Kennedy2012} around eccentric main-sequence star binaries. The existence of these polar circumbinary disks suggests that planet formation in  a polar configuration may be possible. However, to date, all the observed circumbinary planets are close to coplanar to the binary orbital plane \citep[e.g.][]{Doyle2011,Welsh2012,Orosz2012a}. This is likely a result of selection effects \citep{MartinDV2015,MartinDV2017,Czekala2019}. In Section~\ref{polarplanet}, we discuss the evidence for the first polar circumbinary planet. In Section~\ref{implications} we discuss the implications for a polar aligned disk in the AC Her system.   We draw our conclusions in Section~\ref{concs}.

\section{Orientation of the disk relative to the binary orbit}
\label{geo}

\begin{table*}
\begin{center}
\begin{tabular}{l c c c}
\hline
 Parameters of the AC Her system & Symbol & Value & Uncertainty  \\
 \hline
\hline
 \multicolumn{3}{c}{ Binary star parameters from \cite{Anugu2023} }\\
\hline
Mass of post-AGB star & $M_1$ & $0.73\,\rm M_\odot$ & $\pm 0.13\,\rm M_\odot$  \\
Mass of companion star & $M_2$ & $1.4\,\rm M_\odot$ & $\pm 0.12\,\rm M_\odot$  \\
Binary semi-major axis & $a_{\rm b}$ & $2.83\,\rm au$ & $\pm 0.08 \,\rm au$  \\
Binary eccentricity & $e_{\rm b}$ & $0.206 $ & $\pm 0.004$ \\
Binary inclination  &$i_{\rm b}$ & $142.9^\circ$ & $\pm 1.1^\circ$  \\
Binary longitude of ascending node & $\Omega_{\rm b}$ & $155.1^\circ $ & $\pm 1.8^\circ$ \\
Binary argument of periastron & $\omega_{\rm b}$ & $118.6^\circ $ & $\pm 2^\circ$ \\
\hline
 \multicolumn{3}{c}{ Disk parameters from \cite{Hillen2015} }\\
\hline
Disk inclination & $i_{\rm d}$ & $50^\circ $ & $\pm 8^\circ$  \\
Disk longitude of ascending node & $\Omega_{\rm d}$ &  $125^\circ $  & $\pm 10^\circ$\\
\hline
\multicolumn{3}{c}{ Quantities calculated in this work }\\
\hline
Inclination of disk relative to  binary & $i_{\rm bd}$ & 96.5$^\circ$ & \\
Longitude of ascending node of disk relative to  binary & $\Omega_{\rm bd}$ & 84.1$^\circ$ & \\
Inclination of the disk relative to the binary eccentricity vector & $i_{\rm polar}$ & $8.7^\circ$ \\
\hline
\end{tabular}
\end{center}
 \caption{Parameters of the AC Her binary star system. 
}
\label{table2}
\end{table*}

In this section we determine the orientation of the circumbinary disk relative to the orbit of the eccentric binary. The observed binary inclination relative to the plane of the sky, $i_{\rm b}$, longitude of ascending node relative to north, $\Omega_{\rm b}$, and argument of periastron, $\omega_{\rm b}$, are given in Table~\ref{table2}. The observed disk inclination, $i_{\rm d}$ and longitude of ascending node, $\Omega_{\rm d}$ are also given for the disk in Table~\ref{table2}.
The value of  $\Omega_{\rm d}$ is determined by the position angle of the disk as we describe later. 
The position angle has been determined in two independent ways.
\cite{Hillen2015} determined a position angle $125^\circ$ by using mid-infrared Interferometric instrument (MIDI)
on the Very Large Telescope Interferometer (see their Fig. 7).  \cite{Gallardo2021} determined a position angle 
of $136.1^\circ$ on larger scales based on CO emission line velocity information (see their Fig. C.2).
 The disk argument of periastron is not given because it is assumed to be circular.   
 
 We consider a Cartesian coordinate system $(x, y, z)$ 
in which the $x$-axis is along the north direction ($\Omega=0$)  in the plane of the sky and the $z$-axis is along the line of sight towards the observer. The unit binary and disk angular momentum vectors are respectively given by
\begin{align}
    \bm{\hat{l}}_{\rm b} &= (\sin i_{\rm b}\sin \Omega_{\rm b}, -\sin i_{\rm b}\cos \Omega_{\rm b}, \cos i_{\rm b}) \cr
      \bm{\hat{l}}_{\rm d} &= (\sin i_{\rm d}\sin \Omega_{\rm d}, -\sin i_{\rm d}\cos \Omega_{\rm d}, \cos i_{\rm d}), \label{lbd} 
\end{align}
where $\hat{}$ denotes a unit vector.
The unit eccentricity vector of the binary is
\begin{equation}
     \bm{\hat{e}}_{\rm b}= (u_{{\rm b}x}, u_{{\rm b}y},u_{{\rm b}z}) \label{eb}
\end{equation}
 with
\begin{align}
u_{{\rm b}x} & = \cos \omega_{\rm b} \cos\Omega_{\rm b} - \cos i_{\rm b}\sin \omega_{\rm b} \sin \Omega_{\rm b} \cr
u_{{\rm b}y} & =  \cos \omega_{\rm b} \sin \Omega_{\rm b}+ \cos i_{\rm b}\sin \omega_{\rm b}  \cos \Omega_{\rm b}  \cr
u_{{\rm b}z} & = \sin i_{\rm b} \sin \omega_{\rm b}.
\end{align}

Using the binary and disk angular momentum unit vectors given by Equation (\ref{lbd}), we calculate the orientation of the disk relative to the binary. The inclination of the disk relative to the binary is
\begin{equation}
    i_{\rm bd}=\cos^{-1}\left(\bm{\hat{l}_{\rm b}}\cdot \bm{\hat{l}_{\rm d}} \right)
\end{equation}
which agrees  with equation 1 of \cite{Czekala2019}.
The longitude of ascending node of the disk relative to the binary eccentricity vector is 
\begin{equation}
    \Omega_{\rm bd}= \tan^{-1} \left[ \frac{
   \left(\bm{\hat{l}_{\rm b}}\times \bm{\hat{e}_{\rm b}}\right)\cdot \bm{\hat{l}_{\rm d}}}
   {\bm{\hat{l}_{\rm b}}\cdot \bm{\hat{l}_{\rm d}}}
    \right]+\frac{\pi}{2},
\end{equation}
where $\hat{e}_{\rm b}$ is given by Equation (\ref{eb})
\citep[e.g.][]{Chen2019,Chen2020}. 
The misalignment of the disk from polar alignment, or in other words, the inclination of the disk  to the binary eccentricity vector, is
\begin{equation}
    i_{\rm polar}=\cos^{-1}\left( \bm{\hat{l}_{\rm d}}\cdot  \bm{\hat{e}_{\rm b}} \right).
\end{equation}

The possible values of $\Omega_{\rm d}$ determined by the imaging of \cite{Hillen2015} are constrained to lie
along a line at the position angle.  There is then
an ambiguity in the value of $\Omega_{\rm d}$ by a shift of $180^\circ$.
This arises because these observations  cannot distinguish an ascending node
from a descending node.
\cite{Hillen2015} give position angles of $125^\circ$ and $305^\circ$  with $i_{\rm d}=50^\circ$
in both cases. In addition, there is an ambiguity  
in inclination $i_{\rm d}$ that can take on complementary values.
Possible values of $(i_{\rm d}, \Omega_{\rm d})$ are 
$(50^\circ, 125^\circ)$, $(130^\circ, 305^\circ)$, $(50^\circ, 305^\circ)$, and
$(130^\circ, 125^\circ)$. By comparing radiative transfer disk models to data, \cite{Hillen2015}
conclude that the  east side of the disk is farther away (see their Figures 7 and 8).
 This constraint eliminates the latter two configurations.

The degeneracy in the two possible solutions is broken
by using the CO emission line velocity information provided in the studies  
by \cite{Bujarrabal2015} and \cite{Gallardo2021}.
While \cite{Hillen2015} studied the disk out a few hundred au, the velocity studies probed a complementary region that is larger, extending out to about 1000 au. 
In the velocity studies, the disk inclination was found to be $i_{\rm d}=45^\circ$ 
and position angle was determined to be about $135^\circ$. 
The similarity of the position angle to the  \cite{Hillen2015} value suggests that
the disk orientation does not undergo large changes from the inner to outer regions.
Both Fig. 2  of \cite{Bujarrabal2015} and Fig. 2 of \cite{Gallardo2021} provide the line of sight velocity
of the disk as a function of position along the disk equator. These figures show that 
the longitude of the ascending node is on the east side of disk. 
Consequently, only the \cite{Hillen2015}  solution with 
$(i_{\rm d}, \Omega_{\rm d})=(50^\circ, 125^\circ)$ satisfies the velocity constraint.
If we adopt  $(50^\circ, 125^\circ)$, we find $i_{\rm bd}=96.5^\circ$ and $\Omega_{\rm bd}=84.1^\circ$. The disk then is $i_{\rm polar}= 8.7^\circ$ away from polar alignment (polar alignment for a massless disk is $i_{\rm bd}=\Omega_{\rm bd}=90^\circ$). Therefore, the disk is in near polar alignment.

\begin{figure*}
    \centering
    \includegraphics[width=\columnwidth]{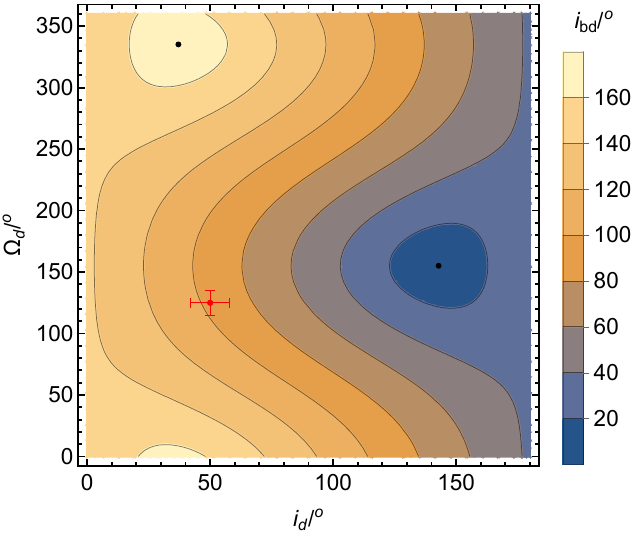}
        \includegraphics[width=\columnwidth]{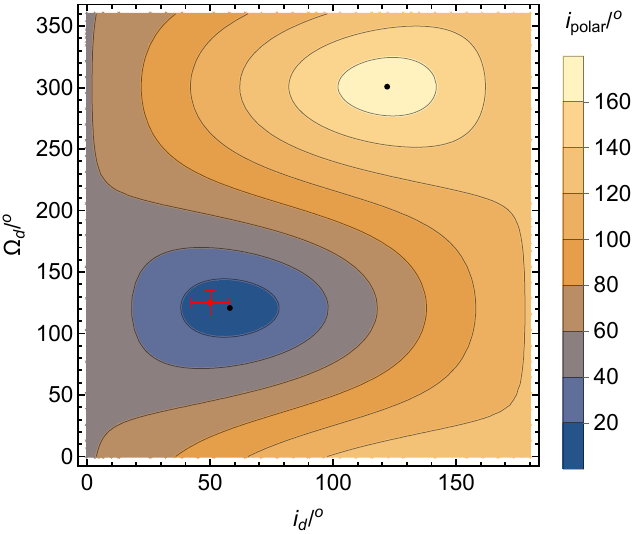}
    \caption{The contours show the inclination of the disk relative to the binary angular momentum vector ($i_{\rm bd}$, left) and relative to the binary eccentricity vector ($i_{\rm polar}$, right) as a function of the observed disk inclination $i_{\rm d}$ and longitude of ascending node $\Omega_{\rm d}$. The red crosses show the possible disk parameters from \cite{Hillen2015}.  The black points on the left panel show a coplanar disk ($i_{\rm bd}=0^\circ$, lower point) and a retrograde coplanar disk ($i_{\rm bd}=180^\circ$, upper point). The black points on the right panel show a polar aligned disk ($i_{\rm bd}=\Omega_{\rm bd}=90^\circ$, lower point) and an anti-polar aligned disk ($i_{\rm bd}=90^\circ$ and $\Omega_{\rm bd}=270^\circ$, upper point). }
    \label{contour}
\end{figure*}

The binary parameters have relatively small uncertainties, however, the disk orientation has larger uncertainties. Therefore, we now consider the range of possible values for $i_{\rm d}$ and $\Omega_{\rm d}$. Fig.~\ref{contour} shows the inclination of the disk away from polar, $i_{\rm polar}$, for the ranges of possible values for the disk angles. A completely polar aligned disk would have $i_{\rm d}=58^\circ$ and $\Omega_{\rm d}=121^\circ$, and this is within the uncertainties of the \cite{Hillen2015} measurements.

\section{First polar planet?}
\label{polarplanet}

The cavity size of a circumbinary gas disk depends upon the binary separation, the binary eccentricity and the inclination of the disk relative to the binary orbit. For a disk that is coplanar to the binary orbit, the disk is truncated in the range $2-3a_{\rm b}$ depending upon the binary eccentricity \citep{Artymowicz1994}. However, this decreases with the inclination of the disk relative to the binary angular momentum vector \citep{Miranda2015,Lubowetal2015,Lubow2018}. The size of the binary cavity is therefore an important diagnostic for the inclination of the disk relative to the binary orbit \citep{Franchini2019inner}. A polar disk may extend down to about $1.6\,a_{\rm b}$.  

The polar configuration of the disk in AC Her does not help to explain the large dust cavity since a polar aligned disk has a smaller cavity size than a coplanar disk. The best explanation remains the presence of a planet in the disk.  Therefore, this is the first evidence of a polar circumbinary planet! Polar circumbinary planets may form as efficiently as coplanar planets \citep[e.g.][]{Childs2021}. A planet in a post-AGB star circumbinary disk could be a second-generation planet, meaning a planet that was formed  from the material of the post-main sequence stellar evolution \citep[e.g.][]{Perets2010, Beuermann2010, Qian2011, Zorotovic2013, Bear2014}. The observational detection limit of $9\,\rm L_\odot$ in \cite{Anugu2023} is inadequate for detecting this low-contrast tertiary. The tertiary's mass is likely low, considering that its gravitational influence remains unobservable in the binary system's orbit. However,  we  do  not attempt to determine the dynamical limit on the tertiary mass in this paper.

In order for dust filtration to occur, the tertiary must  open a gap in the disk \citep[e.g.][]{Zhu2012}. This requires the  planet to be sufficiently massive and in an orbit that is coplanar to the disk, otherwise material can spread across the orbital radius of the planet \citep[e.g.][]{Franchini2020}. While there is no observed dust inside of the orbit of the planet, gas  may still flow inwards past the planet orbit through gas streams \citep{Artymowicz1996,Lubow2006}. Therefore we expect there to be a gas disk interior to the planet orbit unless there are multiple planets interior to the putative planet.  

If the inner circumbinary disk is fed by material flowing inwards from the outer circumbinary disk, then the inner disk must be coplanar to the outer disk, and therefore also polar to the binary orbit. Further, the  timescale for the disk to align to polar is much shorter for the inner disk that is closer to the binary. Since the outer disk is observed to be polar, the inner disk is likely also polar.  Warping of the inner disk may be possible if material is added to the disk at an inclination that is misaligned. Disk warping  generally occurs in a circumbinary disk  that is inclined to the binary orbit. However, a  massless  polar disk has  no  warp. 

The inner gas disk is predicted to be there observationally since the post-AGB star is depleted in refractory elements suggesting that is is accreting material from the circumbinary disk. The accretion of material on to the companion must also be responsible for the formation of the observed jet \citep{Bujarrabal2018,Bollen2020}. We discuss this more in Section~\ref{csd}.

Fig.~\ref{cartoon} shows two visualizations of the system. The disk is composed of two rings that are separated by the gap that the putative planet has carved. The outer disk that is observed is composed of gas and dust, while the inner disk that is not observed is composed only of gas.

\section{Implications of a polar aligned disk}
\label{implications}

We consider here a model in which the disk polar alignment in AC Her results from the evolution of an initially moderately 
misaligned disk to the polar
orientation by means of the tidal evolution of a viscous disk \citep{Aly2015,Martin2017,Lubow2018,Zanazzi2018,Cuello2019,Smallwood2020,Rabago2023}. The application of this model imposes
constraints of the origin and evolution of the circumbinary disk.

\subsection{Formation of the circumbinary disk}

Since all observed post-AGB stars with disks are in binaries, the binary must be an essential component to the disk formation. 
A circumbinary disk that forms misaligned to the binary orbit, evolves towards either coplanar alignment or polar alignment depending upon  its initial inclination \citep{Martin2017}. For a massless disk, the minimum critical inclination relative to the binary orbit is
\begin{equation}
    i_{\rm crit}=\sin^{-1} \sqrt{\frac{1-e_{\rm b}^2}{1+4e_{\rm b}^2}}
\end{equation}
\citep[e.g.][]{Doolin2011}, where $e_{\rm b}$ is the binary eccentricity. For $e_{\rm b}=0.2$, we have $i_{\rm crit}=65^\circ$.  Therefore the disk must be fed material that has, on average, an inclination greater than this if it is to align to polar. If it is fed material with a lower inclination, then it will align towards coplanar. If the binary orbit had a larger eccentricity in the past, then the critical inclination required for a polar disk would have been lower. 

We take the disk age to be of order the post-AGB duration of about $\sim 10^5 \rm yr$ \citep[e.g.,][]{Izzard2023}.
With a precession period of about $\sim 10^4 \rm yr$, a moderately viscous disk can align to polar within its lifetime \citep[e.g.,][]{Martin2017}. 

The nodal precession period for a test particle at orbital radius $R$ that is in a librating orbit is given by
\begin{equation}
    t_{\rm prec}=\frac{2\pi}{\omega_{\rm p}}
\end{equation}
where
\begin{equation}
    \omega_{\rm p}=\frac{3\sqrt{5}}{4}e_{\rm b}\sqrt{1+4e_{\rm b}^2}\frac{M_1M_2}{M_{\rm b}^2} \left(\frac{a_{\rm b}}{R}\right)^{7/2} \Omega_{\rm b}
\end{equation}
\citep{Lubow2018}, where the total binary mass is $M_{\rm b}=M_1+M_2$ and the binary angular frequency is $\Omega_{\rm b}=\sqrt{G M_{\rm b}/a_{\rm b}^3}$. This was derived in the secular quadrupole approximation using equations 2.16-2.18 in \cite{Farago2010}.
For the parameters of AC Her this is
\begin{equation}
     t_{\rm prec}=1.33\times 10^4   \left(\frac{R}{15\,\rm au}\right)^{7/2} \,\rm yr.
\end{equation}
This result suggests that a disk with radius of about $15 \,\rm au$
can evolve to a polar state over
the post-AGB timescale.

A more detailed calculation of a circumbinary disk in which the surface density
varies as $\Sigma \propto 1/R$,
with inner radius $3 a_{\rm b}$ has a precession period of about 
$\sim 2 \times 10^4 \rm yr$ if the  disk outer radius is about $35 \rm au$.
With an  inner disk radius of $1.6 a_{\rm b}$, the same precession period occurs for a  disk  with outer radius  of about  $80 \rm au$.
Such outer radii are smaller than the observed disk outer radius that 
extends to $1000 \rm au$.
Such an extension is possible if the disk viscously expands after
achieving polar alignment, as discussed in the next subsection.

\subsection{Disk Size and Mass}

There is a  discrepancy between the viscous timescale and the disk lifetime for AC Her. An accretion disk spreads out through the effects of viscosity \citep{Pringle1981} that is parameterised with 
\begin{equation}
    \nu=\alpha \left(\frac{H}{R}\right)^2 R^2 \Omega_{\rm K},
\end{equation}
where $\alpha$ is the \cite{SS1973} viscosity parameter, $H/R$ is the disk aspect ratio and $\Omega_{\rm K}=\sqrt{G M_{\rm b}/R^3}$ is the Keplerian angular velocity. 
The viscous timescale in the disk is
\begin{equation}
t_{\nu}=\frac{R^2}{\nu},
\end{equation}
which may be written as 
\begin{equation}
    t_{\nu}=4.5\times 10^3
    \left(\frac{\alpha}{0.1}\right)^{-1}
     \left(\frac{H/R}{0.2}\right)^{-2}
      \left(\frac{R}{30\,\rm au}\right)^{3/2}\,\rm yr.
\end{equation}
The disk aspect ratio for these disks is large \citep[e.g.][]{Kluska2018,Oomen2020} and we choose an appropriate $\alpha$ parameter for a fully ionised disk \citep{Martin2019}.
\cite{Hillen2015} fixed the outer disk radius to $200\,\rm au$ while \cite{Gallardo2021} suggested the outer parts of the disk extend to $1000\,\rm au$. The viscous timescale  at this radius may be of the order of a million  years. This suggests that the disk lifetime must be of this order. However, this is somewhat longer than the disk lifetime suggested by stellar evolution models of up to about $10^5\,\rm yr$ \citep{Izzard2023}.

\cite{Hillen2015} determined the disk dust mass to be greater than $0.001 M_\odot$.  As they point out,
such a large dust mass would imply a very large gas disk mass for the conventional gas/dust ratio of 100 and instead
suggest that lower ratios are more plausible.  The angular momentum of a disk with total mass $0.1\,\rm M_\odot$ would likely not allow for a stable polar disk and instead   von Zeipel-Kozai-Lidov \citep[ZKL,][]{vonZeipel1910,Kozai1962,Lidov1962}  oscillations of the binary would be expected. This has been seen in simulations before in the case of a planet and a star with an outer disk \citep{Terquem2010}. Therefore, we suggest that for AC Her, the total disk angular momentum must be small compared with the binary angular momentum. Indeed, \cite{Gallardo2021} find a total disk mass of $8.1\times 10^{-4}\,\rm M_\odot$  which would allow for a stable polar aligned disk.

\subsection{Feeding of circumstellar disks}
\label{csd}

Simulations  show that material in a  circumbinary disk can flow inwards through the binary cavity and feed the  circumstellar disks around each binary component \citep[e.g.,][]{Artymowicz1996,  Shi2012, DOrazio2013, Munoz2016, Heath2020,Smallwood2023}. Observationally, accretion of material  onto the binary components is inferred in two ways. First, observations show  depletion of refractory elements in the atmospheres of some post-AGB stars. All post-AGB stars that are depleted, including AC  Her, contain a circumbinary disk.  \cite{Oomen2018} suggest that the depletion  is due to the accretion of circumbinary disk gas with low dust content onto the star.  The low dust content is a consequence of the dust trapping described in the Introduction. Second, a jet is often observed from the main-sequence star. The jet in AC Her is tilted by $6.5^\circ$ to the binary angular momentum vector and has a wide opening angle of $30^\circ$ \citep{Bollen2022}. Jets are thought to be perpendicular to the circumstellar accretion disk that powers them \citep{Blandford1982}.

The accretion rate from a polar circumbinary disk  is similar to that of a coplanar circumbinary disk \citep{Smallwood2022}.
A coplanar circumbinary disk feeds the formation of circumstellar disks that are coplanar to the binary orbit. If the circumbinary disk is misaligned to the orbit of the binary, then the circumstellar disks also form misaligned to the binary orbit \citep[e.g.][]{Nixonetal2013}. In the presence of a polar circumbinary disk, the circumstellar disks that form may also form in a polar orientation \citep{Smallwood2023}.

The circumstellar disk undergoes nodal precession about the binary angular momentum vector \citep[e.g.,][]{Larwoodetal1996, Papaloizou1995, lubow2000, Bateetal2000} that, in the presence of dissipation, can lead to coplanar alignment to the binary orbital plane. In addition, for very large misalignments, greater than about $40^\circ$, the circumstellar disks can undergo ZKL oscillations. The inclination and eccentricity of the circumstellar disk can be exchanged. This  can lead on average to a fairly rapid alignment towards the critical ZKL inclination of about $40^\circ$ \citep{Martinetal2014,Fu2015,Martin2016,Lubow2017,Zanazzi2017} but with continuous accretion of high inclination material it can lead to sustained ZKL oscillations \citep{Smallwood2021}.

For a circumstellar disk with a supply of high inclination material, there are competing torques from the addition of material that acts to keep the inclination high and from the binary torque that acts towards coplanar alignment \citep{Smallwood2023}. The outcome depends upon the relative strengths of these torques. If the mass accretion dominates, then the disk inclination can remain high. However, if the binary torque dominates then the disk moves towards coplanar alignment by means of tidal dissipation. The disk also undergoes nodal precession when it is misaligned. Therefore the circumstellar disk could be seen at any inclination and longitude of ascending node. 

The inclination of a jet that is formed by the circumstellar disk around the companion may be perpendicular to the circumstellar disk. Therefore the orientation of a jet being formed around one component of a binary with a  polar circumbinary disk could be in any direction. This is a result of the nodal precession of a misaligned circumstellar disk. This is in contrast to the coplanar circumbinary disk in which the jet orientation would be expected to be close to aligned to the binary angular momentum vector. The large opening angle for the jet in AC Her \citep{Bollen2022} may be indicative of the nodal precession of a circumstellar disk.

\section{Conclusions}
\label{concs}

We have examined the post-AGB star binary system AC Her and shown that it is within $9^\circ$ of polar alignment. In a polar alignment, the disk is inclined by $90^\circ$ to the binary orbital plane with the disk angular momentum vector aligned to the binary eccentricity vector.  This is the first observed polar circumbinary disk around a post-AGB star. 

The inner edge of the dust disk is much farther out than  the tidal truncation radius and a planet within the disk is the most likely explanation. Therefore, this is the first observational evidence of a polar circumbinary planet. While polar gas and debris disks around main-sequence stars have previously been observed, misaligned circumbinary planets have yet to be detected.

The circumbinary disk feeds the formation of circumstellar disks around the binary components. A coplanar circumbinary disk forms coplanar circumstellar disks. However, a polar circumbinary disk forms polar circumstellar disks, but this is not a stable configuration. Polar circumstellar disks may undergo  ZKL inclinations oscillations or 
evolve to coplanarity with the binary. The outcome depends on the competition between the mass accretion torque from the circumbinary disk and the binary torques. Circumstellar disks that are formed from flow from a polar circumbinary disk may be observed at any inclination and orientation.

We describe a model for polar alignment in which the circumbinary disk is initially moderately misaligned with respect to the binary
and evolves to polar alignment by means
of tidal effects. Material from the post-AGB star that is forming the disk 
must have an inclination greater than about $65^\circ$ to the binary orbital plane, assuming the  binary eccentricity
has not evolved significantly. Smaller initial inclinations are possible if the binary eccentricity had been larger in the past.  
Such a model requires the disk to have an outer radius  of less than $100 \rm{au}$, while the disk has been observed out to $\sim 1000 \rm au$ \citep{Bujarrabal2015}.
We suggest that this compact disk expands outwards to reach this larger radius. However, sufficient viscous expansion occurs on timescales of order $10^6 \rm yr$ that is longer than
the duration of the post-AGB phase of about $10^5 \rm yr$.
More analysis should be carried out to better understand the disk properties.

\vspace{0.5cm}
\noindent We thank the anonymous referees for useful comments. RGM and SHL acknowledge support from NASA through grants 80NSSC21K0395 and 80NSSC19K0443. This work is based upon observations obtained with the Georgia State University Center for High Angular Resolution Astronomy Array at Mount Wilson Observatory.  The CHARA Array is supported by the National Science Foundation under Grant No. AST-1636624 and AST-2034336.  Institutional support has been provided from the GSU College of Arts and Sciences and the GSU Office of the Vice President for Research and Economic Development.




\bibliographystyle{aasjournal}
\bibliography{apjl} 



\end{document}